\documentstyle[11pt,aaspp4]{article}

\def\ie{{\it i.e.\hskip 3pt}}

\newcommand{\lya}{Ly$\alpha$\ }
\newcommand{\nh}{N_{\rm HI}}
\newcommand{\kms}{{\rm km}\,{\rm s}^{-1}}
\newcommand{\fn}{f(N_{\rm HI})}
\newcommand\cdunits{{\rm cm}^{-2}}

\slugcomment{Accepted to ApJ}

\lefthead{Dav\'e et al.}
\righthead{Voigt-Profile Analysis of Ly$\alpha$ Forest Simulations}
\singlespace

\begin{document}

\title{Voigt-Profile Analysis of the Lyman-alpha Forest in \\
	a Cold Dark Matter Universe}

\author{Romeel Dav\'e and Lars Hernquist\altaffilmark{1}}
\affil{Astronomy Department, University of California,
    Santa Cruz, CA 95064}

\author{David H. Weinberg}
\affil{Astronomy Department, Ohio State University, Columbus, OH 43210}

\and

\author{Neal Katz}
\affil{Astronomy Department, University of Washington, Seattle, WA 98195}

\altaffiltext{1}{Presidential Faculty Fellow}

\begin{abstract}
We use an automated Voigt-profile fitting procedure to extract statistical
properties of the Ly$\alpha$ forest in a numerical simulation of an $\Omega=1$,
cold dark matter (CDM) universe.  Our analysis method is similar to that 
used in most observational studies of the forest, and we compare the
simulations to recently published results derived from Keck HIRES spectra.
With the Voigt-profile
decomposition analysis, the simulation reproduces the
large number of weak lines ($N_{\rm HI}\la 10^{13}\cdunits$) found
in the HIRES spectra.  The column density distribution evolves significantly
between $z=3$ and $z=2$, with the number of lines at fixed column
density dropping by a factor $\sim 1.6$ in the range where line blending
is not severe.  At $z=3$, the $b$-parameter 
distribution has a median of $35\;\kms$ and a dispersion of $20\;\kms$,
in reasonable agreement with the observed values.
The comparison between our new analysis and recent data strengthens
earlier claims that the \lya forest arises naturally in hierarchical
structure formation as photoionized gas falls
into dark matter potential wells.  However, there are two statistically
signficant discrepancies between the simulated forest and the HIRES
results: the model produces too many lines at $z=3$ by a factor $\sim 1.5-2$,
and it produces more narrow lines ($b<20\;\kms$) than are seen in the data.
The first result is sensitive to our adopted normalization of the mean
\lya optical depth, and the second is sensitive to our assumption
that helium reionization has not significantly raised gas temperatures 
at $z=3$.  It is therefore too early to say whether these discrepancies 
indicate a fundamental problem with the high-redshift structure of the
$\Omega=1$ CDM model or reflect errors of detail in our modeling of the gas 
distribution or the observational procedures.

\end{abstract}


\section{Introduction}

Absorption lines in quasar spectra,
especially the ``forest'' of \lya lines produced by concentrations
of neutral hydrogen, are uniquely suited
for probing structure formation in the high-redshift universe.
The absorbers trace relatively pristine baryonic material over 
a wide range of redshifts, densities, temperatures,
and ionization states.  Recent advances in
computer technology have expanded our ability to predict the
conditions of the absorbing gas,
while high-precision observations made using the HIRES spectrograph 
(\cite{vog94}) on the 10m Keck telescope 
have quantified the statistics of the low column density absorbers to 
unprecedented accuracy (e.g., Hu et al. 1995, hereafter \cite{hu95}).
These data provide stringent
constraints on theories of structure formation and
the state of the intergalactic medium (IGM) at high redshifts.
However, a detailed confrontation between theory and observations
requires that simulations and observed quasar spectra be analyzed
using similar techniques.  In this paper we apply an automated
Voigt-profile fitting algorithm to \lya spectra from a simulation
of the cold dark matter (CDM) scenario (\cite{Pee82}; \cite{Blu84}).
We compare the statistics of the resulting line population to
those derived by \cite{hu95} from HIRES spectra.

Recent cosmological simulations that incorporate gas dynamics, radiative
cooling, and photoionization reproduce many of the observed
features of quasar absorption spectra, suggesting that the \lya forest
arises as a natural consequence of hierarchical structure formation
in a universe with a photoionizing UV background
(\cite{cen94}; \cite{zha95}; Hernquist et al.\ 1996, hereafter
\cite{her96}; \cite{kat96b}; \cite{mir96}).  
Most of the low column density lines are produced by structures
of moderate overdensity that are far from dynamical or thermal 
equilibrium, blurring the traditional distinction between
the \lya forest and Gunn-Peterson (1965) absorption from a 
smooth IGM.
\cite{her96} used a simple flux threshold algorithm to identify
lines in their simulated spectra, defining any region with
transmitted flux continuously below a specified threshold as a 
single line.  They showed that their simulation of a CDM
universe reproduced the observed abundance of absorption systems
as determined by Petitjean et al.\ (1993, hereafter \cite{pet93})
quite well over most of the column density range
$10^{14}\cdunits < \nh < 10^{22}\cdunits$, with a significant
discrepancy for $\nh \sim 10^{17}\cdunits$.

The traditional technique for identifying and characterizing quasar
absorption lines is to fit spectra by a superposition of Voigt profiles.
The HIRES spectra have very high signal-to-noise ratio and resolution,
and most of the lines found in this way are weak absorbers with
column densities $\nh < 10^{14}\cdunits$.  The threshold and Voigt-profile
procedures behave very differently in this regime, since a single feature
identified by the threshold method will often be decomposed into a 
blend of weaker lines when it is modeled as a superposition of Voigt profiles.
In order to compare to published line population statistics from
HIRES data, therefore, it is essential to analyze the simulated
spectra by Voigt-profile decomposition.

The physical model implicit in the decomposition technique
is that of a collection of discrete, compact clouds,
each characterized by a single temperature (or at least by a single
velocity dispersion, which could include contributions from thermal
motion and from Gaussian-distributed ``turbulent'' velocities).
The simulations undermine this physical picture because
the absorbing systems merge continuously
into a smoothly fluctuating background, often contain gas at a range
of temperatures, and are usually broadened in frequency space by
coherent velocity flows that do not resemble Gaussian turbulence.
Nonetheless, any spectrum can be described phenomenologically by
a superposition of Voigt-profile lines, with the number of components
increasing as the signal-to-noise ratio improves and more subtle
features must be matched.  The distributions of fitted column densities
and $b$-parameters provide a useful statistical basis for
comparing simulations and observations, and this is the approach that
we adopt in this paper.  We will discuss the correspondence
between the parameters of the Voigt-profile components and the
physical state of the absorbing gas elsewhere (Dav\'e et al., in preparation).

\section{Simulation and Artificial Spectra}

The simulation analyzed here is the same as that of HKWM:
a CDM universe with $\Omega=1$, 
$H_0 = 50$~km~s$^{-1}$Mpc$^{-1}$, baryon fraction $\Omega_b=0.05$, 
and a periodic simulation cube 22.222 comoving Mpc on a side
containing $64^3$ gas particles and $64^3$ dark matter particles,
with individual particle masses of $1.45\times 10^8 M_\odot$ and
$2.8\times 10^9 M_\odot$, respectively.  
The power spectrum is normalized to $\sigma_8=0.7$, roughly the value
required to reproduce observed galaxy cluster masses (\cite{bah92}; \cite{whi93});
we will consider COBE-normalized CDM models with $\Omega=1$ and $\Omega<1$
in future work.
We use the N-body + smoothed-particle hydrodynamics 
code TreeSPH (\cite{her89}) adapted for
cosmological simulations (Katz, Weinberg, \& Hernquist 1996,
hereafter \cite{kat96}) to evolve the model from $z=49$ to $z=2$.  

Instead of the $\nu^{-1}$ UV background spectrum adopted by HKWM,
we use the spectrum of Haardt \& Madau (1996; hereafter \cite{haa96}),
which is computed as a function of redshift based on the UV output
of observed quasars and reprocessing by the observed \lya forest.
The spectral shape is significantly different from $\nu^{-1}$, but
the UV background influences our \lya forest results primarily
through the HI photoionization rate $\Gamma$, 
a cross-section weighted integral of the spectrum
(\cite{kat96}, equation 29).
The new simulation also includes star formation and feedback 
(see KWH), but this has no noticeable effect on the \lya forest results.  
A comparison between the galaxy populations of this 
simulation and the HKWM simulation appears in
Weinberg, Hernquist, \& Katz (1996).

The mean opacity of the \lya forest depends on the parameter combination
$\Omega_b^2/\Gamma$. 
Since observational determinations of $\Omega_b$ and $\Gamma$
remain quite uncertain, we treat the overall intensity of the UV
background as a free parameter and scale it to match the observed
mean \lya optical depth $\bar\tau_\alpha$.
When evolving the simulation, we divide HM's intensities by a
factor of two, retaining their redshift history and spectral shape.
We find that we must reduce the intensities by further factors of
1.28 and 1.38 at $z=2$ and $z=3$, respectively, in order to match
the estimate $\bar\tau_\alpha = 0.0037(1+z)^{3.46}$ of
Press, Rybicki, \& Schneider (1993; hereafter \cite{pre93}).
Although we apply this final reduction only at the analysis stage,
to compute neutral fractions when generating spectra,
the result is virtually identical to that of changing the
intensity during dynamical evolution (Miralda-Escud\'e et al.\ 1997, 
hereafter \cite{mir97}).  In order to match the PRS mean optical
depth at $z=3$ with the original HM background intensity we would
need $\Omega_b \approx 0.08$, closer to the value advocated by
Tytler, Fan, \& Burles (1996; but see \cite{rug96} and
references therein).
The value of $\bar\tau_\alpha$ plays the role of
a normalizing constraint, used to fix the important combination of free
parameters in our IGM model.  Once $\Omega_b^2/\Gamma$ is set,
there is no further freedom to adjust the simulation predictions,
and the remaining properties of the \lya forest provide tests of
the cosmological scenario itself.

We generate artificial spectra at $z=2$ and $z=3$ along 300 random
lines of sight through the simulation cube, using the
methods described in HKWM and Cen et al.\ (1994).
We do not consider higher redshifts here because the strong absorption
leads to severe blending of lines.
The wavelength spread across the box is 23.4~\AA\ at $z=2$ and
35.9~\AA\ at z=3.
Each artificial spectrum contains 1000 pixels; an individual
pixel has a velocity width $\sim 2\;\kms$ and a spatial extent
$\sim 20$ comoving kpc, twice the gravitational softening length.
In the \lya forest regime, the gas distribution is smooth on these scales.

\section{Fitting Voigt Profiles to Artificial Spectra}

We want the analysis of our simulated spectra to closely match
that used in typical observational studies, HKCSR in particular.
To this end, we have developed an automated Voigt-profile fitting routine,
AUTOVP, which allows us to efficiently handle large quantities of
simulated data and which provides an objective algorithm that
can be applied to observational data.

We add noise to our simulated spectra employing a combination of
Gaussian photon noise with signal-to-noise ratio $S/N=50$
in the continuum (corresponding roughly to the \cite{hu95} data)
and a fixed readout noise chosen to match the 
characteristics of the Keck HIRES spectrograph.
Varying $S/N$ changes our results only at the lowest column densities.
While we know the true continuum level in the 
simulated spectra, this is not the case for the observational data.
We therefore estimate the continuum in the simulated spectra by
the iterative procedure commonly used for Echelle data:
fitting a third-order polynomial to the
data set, excluding any points lying $\ga 2\sigma$
below the fit, refitting the non-excluded points, and repeating
until convergence is achieved.  
This technique is based upon the implicit assumption that the regions of
lowest absorption in a high-resolution spectrum lie close to the
true continuum level.
Because the simulated spectra show 
fluctuating Gunn-Peterson absorption that increases with $z$,
continuum fitting has a systematic tendency to remove flux,
an average of 1.2\% at $z=2$ and 5.7\% at $z=3$.
The effect would probably be somewhat smaller in observational data
because a typical HIRES Echelle order ($\sim 45$\AA) is longer than
one of our simulated spectra, giving a higher probability that
the spectrum contains a region of genuinely low absorption.  

Given a normalized, continuum-fitted spectrum and its noise vector,
we apply AUTOVP to detect lines and fit Voigt profiles.  
In its first phase, AUTOVP identifies lines and makes an initial
estimate of their column densities and $b$-parameters.  
Detection regions are identified above an 8$\sigma$ confidence
level, following the method of \cite{lan87}.
For line identification purposes, the data is convolved with a two-pixel-width
Gaussian and $1\sigma$ of noise is subtracted to yield a
``minimum flux".
For non-saturated regions, a single Voigt profile is placed at the 
lowest flux value in the detection region, 
and $N_{\rm HI}$ and $b$ are 
reduced by small increments from large initial values until the model is
everywhere above the minimum flux.  
For saturated regions a line is placed in the middle of the trough, and 
$N_{\rm HI}$ and $b$ are adjusted to fit the ``cusp" regions, \ie
regions about five pixels wide on either side of the trough.
The resulting first-guess line 
is then subtracted from the data to obtain a residual flux.  
Detections regions are identified in
the residual flux, and the procedure is repeated until there are no more
$8\sigma$ detections.
The line identification
procedure is very robust, never failing for non-saturated
regions and only occasionally producing a bad fit even in complex,
blended, saturated regions, where $N_{\rm HI}$ and $b$ are largely degenerate
and the cusp regions are difficult to identify.

In its second phase, AUTOVP takes the initial guess 
and performs a simultaneous $\chi^2$-minimization on the parameters
($v_{\rm central}, N_{\rm HI}, b$) of all lines within each detection region.  
Three independent minimization techniques are employed in conjunction
in order to reliably identify the global $\chi^2$ minimum.
AUTOVP then tries to remove any components 
with a formal error in $N_{\rm HI}$ or
$b$ comparable to the parameter value, refitting
the detection region with one less line.  If the resulting
$\chi^2$ is lower than the original value the rejection is accepted,
otherwise the fit returns to the original set of lines.  
AUTOVP thus attempts to fit the spectrum with as few lines as possible
while still minimizing $\chi^2$.  If the fit after these line rejections
is ``good" (characterized empirically by $\chi^2 \la 2$ per pixel), 
the program ends, 
otherwise it tries to add a line where the local contribution to
$\chi^2$ is greatest.  
The rare cases where AUTOVP fails to find a good fit
are flagged for possible manual intervention.
AUTOVP is designed to interface with the PROFIT
interactive Voigt-profile fitting package (\cite{chu96}).  
This graphical interactive fitter can be used to 
manually adjust poor fits, although this was required in so few cases
that we base our simulation statistics entirely on the automated fits.

\placefigure{fig: autovp_ex}

In Figure~\ref{fig: autovp_ex} we show the results of AUTOVP applied
to a $S/N=50$, continuum-fitted spectrum at $z=3$.
The bottom panel shows the 
first-guess fit superimposed on the simulated spectrum.
The top panel shows the final fit after the
$\chi^2$-minimization has been performed.
Generally AUTOVP has greatest difficulty in 
obtaining first-guess fits in blended saturated regions like the one
illustrated here.  Nevertheless, the minimization 
produced an adequate fit with no interactive adjustment.
Lines with $N_{\rm HI} \geq 10^{13}$ are indicated by the
long tick marks above the spectrum, while lines with 
$N_{\rm HI} < 10^{13}$ have short tick marks.
The number of these small lines identified by AUTOVP is somewhat
sensitive to the adopted $S/N$ and the detection threshold.
To keep our results fairly robust against details of our fitting procedure,
we exclude these lines from our analysis and only focus
on lines with $N_{\rm HI} \geq 10^{13}$.
AUTOVP has also been applied to observational data from H1216 and MgII2796,
with results quite similar to those obtained from manual fitting.

\section{Results}

\placefigure{fig: col}

Figure~\ref{fig: col} shows the column density distribution 
$\fn$, the number of lines per unit
redshift per linear interval of $\nh$.
Solid and dashed lines show the simulation results from AUTOVP
at $z=3$ and $z=2$, respectively.  The dotted line shows $\fn$
obtained using the HKWM threshold algorithm at $z=3$, with a
flux threshold of 0.7.  As expected, the two methods yield
similar results at high column densities, $\nh \ga 10^{14.5}\;\cdunits$,
but at lower $\nh$ AUTOVP deblends much more and finds many more lines.
We find a similar trend at $z=2$, though because of the reduced line
crowding at lower redshift the agreement between the two methods
extends down to $\nh \sim 10^{14}\;\cdunits$.

Filled and open circles in Figure~\ref{fig: col} show the observational
results of PWRCL and HKCSR, respectively.
The two determinations of $\fn$ agree well in their regime of overlap,
$10^{13.6}\;\cdunits \la \nh \la 10^{14.3}\;\cdunits$.
The high $S/N$ and resolution of the HIRES data allow HKCSR to
detect much weaker absorption features, and their $\fn$ continues
to rise down to the lowest bin, $\nh \sim 10^{12.5}\;\cdunits$.
Clearly a comparison between the simulations and HKCSR's published
line statistics must be based on Voigt-profile fitting, since
the vast majority of their lines lie in the region where line blending
causes large differences between this method and the threshold algorithm.
We compute the HKCSR $\fn$ directly from their published line list,
with no corrections for ``incompleteness.''  HKCSR estimate such
corrections from artificial spectra {\it assuming} an underlying model
of randomly distributed, Voigt-profile lines with a power-law $\fn$
and a specified distribution of $b$-parameters, and they conclude
that their results are consistent with $\fn \propto \nh^{-1.46}$ down
to $\nh \approx 10^{12.5}\;\cdunits$, where the correction for
incompleteness is a factor of four.  If we applied the same correction
factors to the simulation results, the derived column density 
distributions would also rise in a nearly power-law fashion instead
of turning over at low column densities.  But the simulations provide
no a priori reason to expect Voigt-profile lines or a power-law $\fn$,
so we prefer to compare them directly to the data without trying to
correct either for lines ``lost'' to blending.

The mean redshift of the HKCSR lines is $\bar z=2.9$, so the
closest comparison is to the $z=3$ simulation results.
To make this comparison more exact, we convolved the $z=3$ artificial
spectra to a resolution
$\Delta\lambda = 0.06$\AA\ ($\Delta v=3.7\;\kms$) before analysis,
which has the minor effect of removing some lines with
$N_{\rm HI}\la 10^{13}$.
When analyzed by Voigt-profile decomposition, the simulation reproduces
the large number of weak lines found in the HIRES spectra.  In fact,
the simulation overproduces the number of lines by a factor of $1.5-2$
in the column density range
$10^{13}\;\cdunits\la N_{\rm HI}\la 10^{14}\;\cdunits$,
a discrepancy that we will return to in \S 5.
The rolloff at low column densities is also somewhat different,
but the results for the weakest lines are the most sensitive
to the details of the fitting procedure and to the modeling
of noise and spectral resolution, so we regard this difference 
as less significant.

In the regime where line blending is unimportant, $\fn$ of the
simulations drops by a factor of $\sim 1.5-2$ between $z=3$ and $z=2$.
At low column densities $\fn$ actually increases because of the
reduced effects of line blending.
As discussed in HKWM, Miralda-Escud\'e et al.\ (1996), 
and \cite{mir97}, the evolution
of the line population over this redshift range is driven primarily
by the expansion of the universe, which lowers the physical gas
densities in the absorbing systems and thereby lowers their neutral
fractions and corresponding HI column densities.  
It is therefore more physically appropriate to think of $\fn$
as evolving to the left rather than evolving downwards, though the
quantitative effect is the same to within the accuracy of this
simplified account.

\placefigure{fig: bpar}

Figure~\ref{fig: bpar} shows the distribution of $b$-parameters
for lines with $\nh \geq 10^{13}\;\cdunits$ from HKCSR
(solid histogram) and from the AUTOVP analyses of the simulation at $z=3$
and $z=2$ (solid and dashed curves, respectively).
The $10^{13}\;\cdunits$ cutoff eliminates lines whose identification
and derived properties are sensitive to the value of $S/N$ or to
details of the fitting procedure, though the results 
do not change qualitatively if we lower this cutoff to 
$10^{12.5}\;\cdunits$.
The threshold method (dotted curve) 
yields much larger $b$-parameters than AUTOVP at $z=3$
because many of its identified ``lines'' are 
extended absorption regions, which AUTOVP separates into narrower components.
Table~\ref{table: bdist} lists the median, mean, and $1\sigma$
dispersion of the $b$-parameter histograms.  The $z=3$ simulation 
values for all three numbers are slightly larger than the HKCSR values,
but the agreement is quite good given that the analysis procedures
are not identical in all their details.  The most significant difference
in the distributions is the presence of many more narrow ($b<20\;\kms$)
lines in the simulation than in the data, a discrepancy that we
discuss further below.

\section{Discussion}

Our most important result is that the CDM simulation reproduces
the large number of weak lines found in HIRES spectra when it
is analyzed by Voigt-profile decomposition.
However, in the column density range
$10^{13}\;\cdunits \la N_{\rm HI} \la 10^{14}\;\cdunits$,
the density of lines in the simulation at $z=3$ is a factor of 1.5--2
higher than found by HKCSR at $\bar z=2.9$.
Our simulation suffers from the inevitable limitation of finite
numerical resolution, but the \lya absorbers are usually large, smooth,
low-overdensity structures, and we would in any case expect higher
numerical resolution to increase the number of lines rather than 
decrease it.  This excess of lines may therefore indicate a failure
of the $\Omega=1$, $\sigma_8=0.7$ CDM model, a tendency to produce
too much small scale clumping at $z=3$.

An alternative possibility, quite plausible at present, is that
we have set the intensity of the UV background too low given our
adopted value of $\Omega_b$.  As discussed in \S 2, we choose the
background intensity in order to match the PRS determination 
of the mean \lya optical depth, $\bar\tau_\alpha = 0.45$
at $z=3$.  The statistical uncertainties in this determination are
fairly small, but there are systematic uncertainties in the required
extrapolation of the quasar continuum into the \lya forest region.
We can match the HKCSR results if we
increase the UV background intensity by a factor $\sim 1.7$,
thus lowering HI column densities by a similar factor and shifting
the simulation result for $\fn$ to the left.
This increase lowers the mean optical depth to $\bar\tau_\alpha=0.32$,
which is well outside the $1\sigma$ range of PRS (figure~4) at
the HKCRS mean redshift $\bar z=2.9$ but is consistent with the PRS 
value at $z=2.65$.
It is somewhat {\it above} the value 
$\bar\tau_\alpha(z=3) \sim 0.25$ found by \cite{zuo93},
who use a different data set and a different
method of determining the quasar continuum.
The uncertainty of our conclusions highlights the need for better
observational determinations of $\bar\tau_\alpha(z)$,
which plays a crucial role in normalizing the predictions of
cosmological simulations.
The mean optical depth depends on the distribution of $b$-parameters
and on the small scale clustering of absorbers in addition to the
column density distribution itself, so if $\bar\tau_\alpha$ is
well known then the amplitude of $\fn$ becomes an important 
independent test of the high-redshift structure predicted by
a cosmological model.

A second important result of our comparison is that the 
$\Omega=1$, $\sigma_8=0.7$ CDM model produces \lya forest lines
with typical $b$-parameters close to observed values.
However, the simulation yields many more
lines with $b<20\;\kms$ than are found by HKCSR.
A thermally broadened, single-temperature gas cloud produces
a \lya absorption line of width $b=20(T/24,000\;{\rm K})^{1/2}\;\kms$.
Low-$b$ lines arise in the simulated spectra because much
of the absorbing gas is at temperatures of $10^4\;$K or less,
with its temperature determined by the balance between
photoionization heating and adiabatic cooling (\cite{mir97}).
We will need tests at higher resolution to check that these
temperatures are not artificially low because the simulation
misses entropy production in unresolved shocks, but because such
shocks would have to be quite weak, we expect that the effect
of missing them would be small.

Miralda-Escud\'e \& Rees (1994) 
pointed out that the process of reionization can
heat the IGM significantly if it occurs fast enough to prevent
radiative cooling losses.
Our equilibrium treatment of photoionization (KWH) implicitly
suppresses this effect, and since 
circumstantial evidence suggests that HeII reionization 
may have occurred at $z \approx 3$ (\cite{son96}), we could be
underestimating the gas temperatures at this redshift.
The dot-dash line of Figure~\ref{fig: bpar} shows the $b$-parameter
distribution obtained at $z=3$ after adding 15,000 K to the temperatures
of the SPH particles (a thermal energy equivalent to 4 Rydbergs per
HeII photoelectron), reducing the UV background intensity by 2.46
to restore $\bar\tau_\alpha=0.45$, then reextracting and reanalyzing spectra.
Heating the gas eliminates the excess of low $b$-parameter lines,
though it worsens the agreement with \cite{hu95} at $b>40\;\kms$.
We will investigate other treatments of reionization heating
in future work, though the effects will be difficult to pin down
because they depend on uncertain details of reionization (\cite{mir94}).
Model predictions for $b$-parameters should be more robust at $z=2$,
since by this time much of the energy absorbed during HeII
reionization at $z\ga 3$ will have been lost to adiabatic cooling.

Sharper tests of cosmological models against the statistics of the \lya
forest can be obtained by expanding the redshift range of comparisons,
by improving the determination of $\bar\tau_\alpha(z)$, and by
applying AUTOVP to observational data, so that the analyses of simulated
and observed spectra are identical in detail.
We will also test models of the \lya forest using alternative statistical
measures to characterize spectra, for if the physical scenario that
emerges from cosmological simulations is correct, then
Voigt-profile decomposition provides at best a rough guide
to the density and temperature profiles of the absorbing gas.
At the high $S/N$ and resolution of the HIRES data, an
absorption feature with a single flux minimum 
often shows asymmetries or broad wings, requiring two or more Voigt-profile
lines to provide an adequate fit (\cite{hu95}).
Pairs of lines with small velocity separations have strongly
anti-correlated $b$-parameters, suggesting that many of these decompositions
are not genuine physical blends (\cite{rau96}).
While results such as these can always be accommodated within a 
discrete ``cloud'' model by postulating just the right
clustering properties, they more likely signify the breakdown of
the Voigt-profile paradigm itself, revealing the origin of the \lya
forest in the diffuse, undulating gas distribution
of the high-redshift universe.

\acknowledgments
We acknowledge the invaluable assistance of Chris Churchill and
numerous stimulating discussions with Jordi Miralda-Escud\'e.  We thank 
the authors of \cite{hu95} for making their line list available.
We also thank Renyue Cen for his timely refereeing and helpful comments.
This work was supported in part by the PSC, NCSA, and SDSC supercomputing
centers, by NASA theory grants NAGW-2422, NAGW-2523, NAG5-2882, and NAG5-3111,
by NASA HPCC/ESS grant NAG 5-2213,
and by the NSF under grants ATS90-18256, ASC 93-18185 and the Presidential
Faculty Fellows Program.  


%
%

\clearpage


%
%

\clearpage

\figcaption[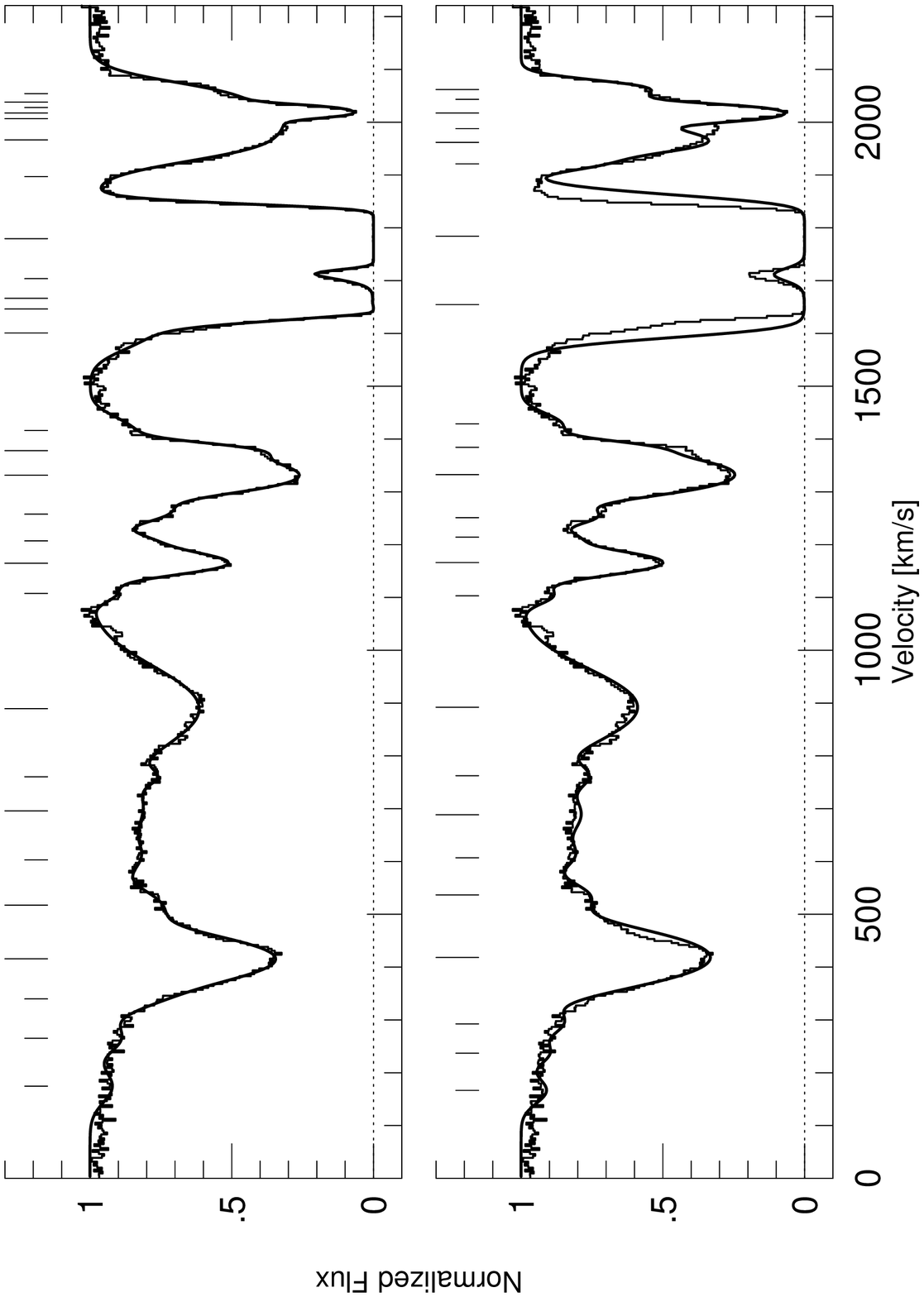]{Example AUTOVP fit (smooth curves) for a 
continuum-normalized, $z=3$, artificial spectrum (histograms).
The bottom panel shows the first-guess fit, while the
top panel shows the final fit after $\chi^2$-minimization.
Long tick marks indicate lines with $N_{\rm HI} \geq 10^{13}$, 
while short tick marks indicate lines with $N_{\rm HI} < 10^{13}$.
\label{fig: autovp_ex}}

\figcaption[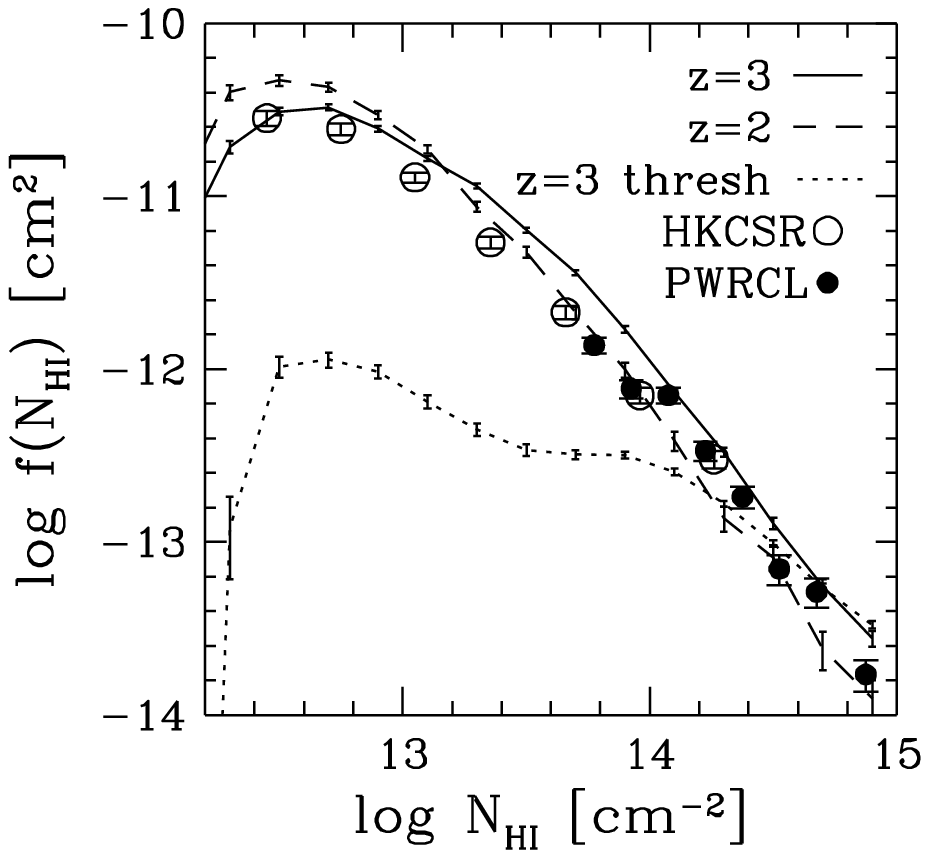]{
Column-density distributions $\fn\equiv d^2N/dz\,dN_{\rm HI}$, the number
of lines per unit redshift per linear interval of HI column density.
Solid and dashed lines show the AUTOVP results for 300 spectra along
random lines of sight through the simulation at $z=3$ and $z=2$, respectively.
The dotted line shows $\fn$ obtained by the threshold method at $z=3$.
Filled and open circles show the observational results of PWRCL and
HKCSR, respectively.  Error bars denote $1\sigma$ Poisson counting errors.
\label{fig: col}}

\figcaption[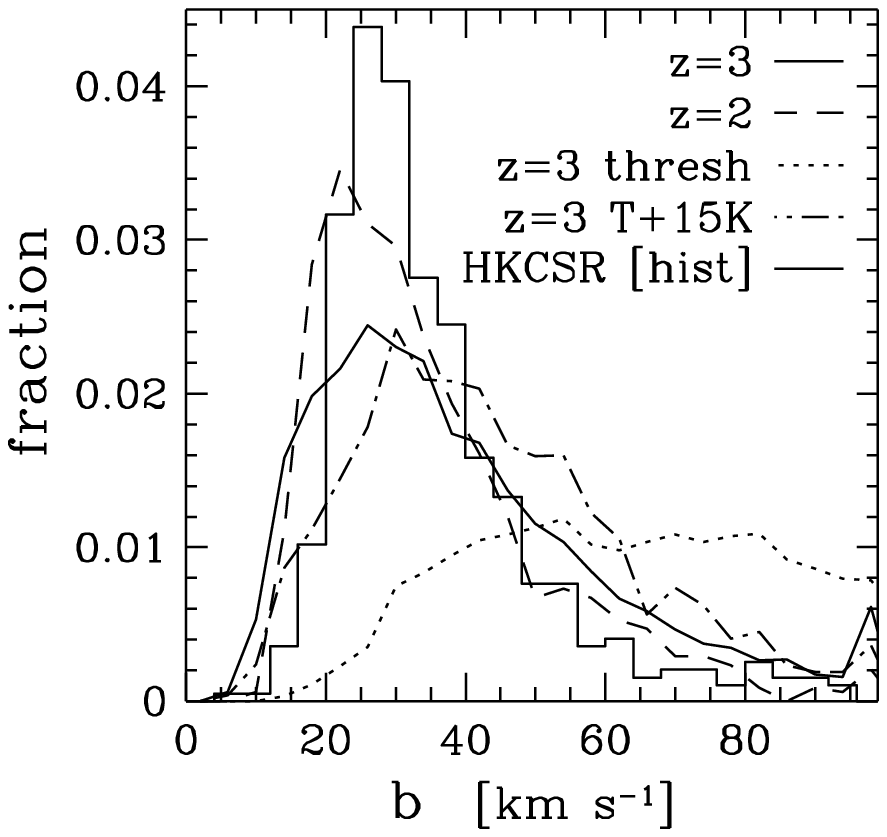]{
Distributions of $b$-parameters, for lines with 
$N_{\rm HI}>10^{13}\cdunits$.  Solid and dashed curves show the AUTOVP
results from the simulation at $z=3$ and $z=2$, respectively.  The dotted
line shows the result of the threshold method at $z=3$;
blending of features leads to very large line widths.
The dot-dash line shows the AUTOVP result at $z=3$ after gas temperatures
have been increased by $15,000$ K.  The solid histogram shows the
HKCSR data.  All distributions are computed as histograms with the same
bins, but the simulation results are shown as curves to prevent visual
confusion.
\label{fig: bpar}}


\clearpage
 
\begin{deluxetable}{cccccc}
\footnotesize
\tablecaption{Moments of $b$-parameter distributions. \label{table: bdist}}
\tablewidth{0pt}
\tablehead{
\colhead{Spectra} & \colhead{Algorithm} & \colhead{$\bar{z}$} & \colhead{$b_{\rm med}$} & \colhead{$b_{\rm mean}$} &\colhead{$\sigma_b$}
}
\startdata
HKCSR   &       &2.9 &31.4 &35.8 &16.3\nl
CDM  &AUTOVP &3   &34.6 &39.3 &20.8\nl
CDM  &AUTOVP &2   &30.6 &34.7 &16.5\nl
CDM+15K &AUTOVP &3 &40.9 &43.8 &19.5\nl
CDM  &Threshold &3   &74.5 &83.9 &47.6\nl
CDM  &Threshold &2   &43.4 &49.9 &26.3\nl
 
\enddata
\end{deluxetable}

\clearpage

\plotone{fig.auto9.ps}

\clearpage

\centerline{
\epsfxsize=3.5truein
\epsfbox[45 470 315 725]{fig.col.ps}
}



\centerline{
\epsfxsize=3.5truein
\epsfbox[50 470 310 725]{fig.bpar.ps}
}


\begin{thebibliography}{}

\bibitem[Bahcall \& Cen 1992]{bah92} Bahcall, J. \& Cen, R. 1992
    \apjl, 398, L81
\bibitem[Blumenthal et al.\ 1984]{Blu84} Blumenthal, G. R., Faber, S. M., 
	Primack, J. R., \& Rees, M. J. 1984, Nature, 311, 517
\bibitem[Cen et al. 1994]{cen94} Cen, R., Miralda-Escud\'e, J.,
    Ostriker, J.P., \& Rauch M. 1994, \apj, 427, L9
\bibitem[Churchill 1996]{chu96} Churchill, C.W. 1996, UCSC Ph.D. Thesis
\bibitem[Gunn \& Peterson 1965]{gun65} Gunn, J.E. \& Peterson, B.A. 1965,
    \apj, 142, 1633
\bibitem[HM]{haa96} Haardt, F. \& Madau, P. 1996, \apj, 461, 20 (\cite{haa96})
\bibitem[Hernquist \& Katz 1989]{her89} Hernquist, L.H. \& 
    Katz, N. 1989, \apjs, 70, 419
\bibitem[HKWM]{her96} Hernquist, L.H., Katz, N., Weinberg, D.H., 
    \& Miralda-Escud\'e, J. 1996, \apjl, 457, L51 (\cite{her96})
\bibitem[HKCSR]{hu95} Hu, E.M., Kim, T.S., Cowie, L.L.,
    Songaila, A., \& Rauch, M. 1995 \aj, 110, 1526 (\cite{hu95})
\bibitem[KWH]{kat96} Katz, N., Weinberg D.H., \& Hernquist, L. 1996, 
    \apjs, 105, 19 (\cite{kat96})
\bibitem[Katz et al.\ 1996b]{kat96b} Katz, N., Weinberg D.H., Hernquist, L.,
    \& Miralda-Escud\'e, J. 1996b, \apjl, 457, L57
\bibitem[Lanzetta, Turnshek, \& Wolfe (1987)]{lan87} 
    Lanzetta, K.M., Turnshek D.A., \& Wolfe, A.M. 1987, \apj, 322, 739
\bibitem[Miralda-Escud\'e et al. 1996]{mir96} Miralda-Escud\'e, J.,
    Cen, R., Ostriker, J.P., \& Rauch, M. 1996, \apj, in press,
    astro-ph/9511013
\bibitem[Miralda-Escud\'e \& Rees 1994]{mir94} Miralda-Escud\'e, J. \&
    Rees, M. 1994, \mnras, 266, 343
\bibitem[MWHK]{mir97} Miralda-Escud\'e, J., Weinberg, D. H., Hernquist, L.,
    \& Katz, N. 1997, in preparation (\cite{mir97})
\bibitem[Peebles 1982]{Pee82} Peebles, P. J. E. 1982, \apjl, 263, L1
\bibitem[PWRCL]{pet93} Petitjean, P., Webb, J.K.,
    Rauch, M., Carswell, R.F, \& Lanzetta, K. 1993, \mnras, 262, 499 
    (\cite{pet93})
\bibitem[Press \& Rybicki 1993]{pr93} Press, W.H. \& Rybicki, G.B.
    1993, \apj, 418, 585
\bibitem[PRS]{pre93} Press, W.H., Rybicki, G.B., \&
    Schneider, D.P. 1993, \apj, 414, 64 (\cite{pre93})
\bibitem[Rauch 1996]{rau96} Rauch, M. 1996, in Cold Gas at High Redshift,
    eds. M. Bremer, H. Rottgering, C. Carilli, \& P. van de Werf,
    (Dordrecht: Kluwer)
\bibitem[Rugers \& Hogan 1996]{rug96} Rugers, M. \& Hogan, C. J. 1996,
    ApJ, 459, L1
\bibitem[Songaila \& Cowie 1996]{son96} Songaila, A. \& Cowie, L. L. 1996,
    AJ, 112, 335
\bibitem[Tytler, Fan, \& Burles 1996]{tyt96} Tytler, D., Fan, X.M.,
    \& Burles, S. 1996, Nature, 381, 207
\bibitem[Vogt, et al. 1994]{vog94} Vogt, S. S., et al. 1994, SPIE, 2198, 326
\bibitem[Weinberg et al.\ 1996]{wei96} Weinberg, D. H., Hernquist, L.,
    \& Katz, N. 1996, ApJ, submitted, astro-ph/9604175
\bibitem[White, Efstathiou, \& Frenk 1993]{whi93} White, S. D. M.,
    Efstathiou, G., \& Frenk, C. S. 1993, \mnras, 262, 1023
\bibitem[Zhang, Anninos, \& Norman 1995]{zha95} Zhang, Y., Anninos, P.,
    \& Norman, M.L. 1995, \apjl, 453, L57
\bibitem[Zuo \& Lu (1993)]{zuo93} Zuo, L. \& Lu, L. 1993, \apj, 418, 601

\end{thebibliography}
\end{document}